\documentclass[aps,pra,twocolumn,superscriptaddress,10pt]{revtex4-1}

\usepackage[english]{babel}
\usepackage{amsmath}
\usepackage{amsthm}
\usepackage{amssymb}
\usepackage{mathrsfs}
\usepackage{booktabs}
\usepackage{color}
\usepackage{hyperref}
\usepackage[ruled,vlined]{algorithm2e}
\usepackage{cases}
\usepackage{enumitem}

\usepackage{multirow}
\hypersetup{
  colorlinks   = true,  
  urlcolor     = blue,  
  linkcolor    = blue,  
  citecolor    = red    
}
\usepackage{mathtools}
\usepackage{natbib}
\usepackage{pgfplots}
\pgfplotsset{compat=1.18}
\usepackage{bbm}
\usepackage{subfigure}
\usepackage{url}
\usepackage{tabularx}
\newcolumntype{P}[1]{>{\centering\arraybackslash}p{#1}}

\newcommand{\di}{{\rm d}}

\mathtoolsset{showonlyrefs}

\theoremstyle{definition}

\theoremstyle{remark}

\theoremstyle{plain}
\newtheorem*{theorem*}{Theorem}

\usepackage{scalerel,stackengine}
\stackMath
\newcommand\reallywidehat[1]{%
\savestack{\tmpbox}{\stretchto{%
  \scaleto{%
    \scalerel*[\wi\di thof{\ensuremath{#1}}]{\kern-.6pt\bigwedge\kern-.6pt}%
    {\rule[-\textheight/2]{1ex}{\textheight}}
  }{\textheight}%
}{0.5ex}}%
\stackon[1pt]{#1}{\tmpbox}%
}
\usepackage{comment}
\usepackage{float}

\SetArgSty{textnormal}

\makeatletter

\newenvironment{widetext2}{%
  \par\ignorespaces
  \setbox\widetext@top\vbox{%
   \vskip15\p@
   \hb@xt@\hsize{%
    \leaders\hrule\hfil
    \vrule\@height6\p@
   }%
   \vskip6\p@
  }%
  \setbox\widetext@bot\hb@xt@\hsize{%
    \vrule\@depth6\p@
    \leaders\hrule\hfil
  }%
  \onecolumngrid
  \let\set@footnotewidth\set@footnotewidth@ii
}{%
  \par
  \twocolumngrid\global\@ignoretrue
  \@endpetrue
}%

\makeatother

\begin{document}

\title{Consensus-based qubit configuration optimization for variational algorithms on neutral atom quantum systems}

\author{R.J.P.T. \surname{de Keijzer}}
\altaffiliation[Corresponding author: ]{r.j.p.t.d.keijzer@tue.nl }
\affiliation{
Department of Applied Physics and Science Education, Eindhoven University of Technology, P.\ O.\ Box 513, 5600 MB Eindhoven, The Netherlands}
\affiliation{
Eindhoven Hendrik Casimir Institute, Eindhoven University of Technology, P.\ O.\ Box 513, 5600 MB Eindhoven, The Netherlands
}

\author{L. Y. \surname{Visser}}
\affiliation{
Department of Mathematics and Computer Science, Eindhoven University of Technology, P.\ O.\ Box 513, 5600 MB Eindhoven, The Netherlands
}
\affiliation{
Eindhoven Hendrik Casimir Institute, Eindhoven University of Technology, P.\ O.\ Box 513, 5600 MB Eindhoven, The Netherlands
}

\author{O. \surname{Tse}}
\affiliation{
Department of Mathematics and Computer Science, Eindhoven University of Technology, P.\ O.\ Box 513, 5600 MB Eindhoven, The Netherlands
}
\affiliation{
Eindhoven Hendrik Casimir Institute, Eindhoven University of Technology, P.\ O.\ Box 513, 5600 MB Eindhoven, The Netherlands
}

\author{S.J.J.M.F. \surname{Kokkelmans}}
\affiliation{
Department of Applied Physics and Science Education, Eindhoven University of Technology, P.\ O.\ Box 513, 5600 MB Eindhoven, The Netherlands}
\affiliation{
Eindhoven Hendrik Casimir Institute, Eindhoven University of Technology, P.\ O.\ Box 513, 5600 MB Eindhoven, The Netherlands
}

\date{\today}

\begin{abstract}
In this work, we report an algorithm that is able to tailor qubit interactions for individual variational quantum algorithm problems. Here, the algorithm leverages the unique ability of a neutral atom tweezer platform to realize arbitrary qubit position configurations \cite{madjarov}. These configurations determine the degree of entanglement available to a variational quantum algorithm via the interatomic interactions. Good configurations will accelerate pulse optimization convergence and help mitigate barren plateaus. As gradient-based approaches are ineffective for position optimization due to the divergent $R^{-6}$ nature of neutral atom interactions, we opt to use a \textit{consensus-based} algorithm to optimize the qubit positions. By sampling the configuration space instead of using gradient information, the consensus-based algorithm is able to successfully optimize the positions, yielding adapted variational quantum algorithm ansatzes that lead to both faster convergence and lower errors. In this work, we show that these optimized configurations generally result in large improvements in the system's ability to solve ground state minimization problems for both random Hamiltonians and small molecules.
\end{abstract}

\maketitle

\section*{INTRODUCTION}
\label{sec:introduction}

The goal of a variational quantum algorithm (VQA) is to construct a parametrized unitary $U$ that maps an initial state $|\psi_0\rangle$ to a final state $|\psi(T)\rangle$ minimizing some cost function $f$ \cite{vqas}. This is performed by optimizing the parameters $\theta$ in parameterized unitaries $U[\theta]$, such that $|\psi(T)\rangle=U[\theta]|\psi_0\rangle$. Generally, gate-based variational quantum algorithms will try to create a universal gate ansatz that is in theory able to find a minimizer for any $f$ \cite{nielsen2010quantum}. Such ansatzes include the hardware-efficient ansatz \cite{hardwareefficient} and the qubit coupled cluster ansatz (QCC) \cite{qcc}.\\
 
 However, these ansatzes often require a multitude of gates and large depths to realize specific unitaries. Especially in the NISQ era \cite{Preskill_2018}, these large-depth circuits can lead to low fidelities that inhibit the rendering of the unitary. Recent work suggests using a problem-inspired ansatz rather than a universal one \cite{adapt-vqe,adapt2,adapt3,uccsd}. This could lead to lower depths and faster convergence in finding optimal parameters $\theta$. In quantum chemistry, VQAs are used to find the ground state of some molecular target Hamiltonian $H_{\text{targ}}$ by minimizing a function of the form $f=\langle\psi(T)|H_{\text{targ}}|\psi(T)\rangle$. Work on problem-inspired ansatzes for these problems includes the UCCSD ansatz based on the annihilation and creation operators of electronic orbitals \cite{uccsd}, and the ADAPT-VQE ansatz which tries to gain the most correlation energy for the least number of parameters \cite{adapt-vqe}. We refer to  Ref.~\cite{ansatzoverview} for a recent comprehensive overview on gate-based problem-inspired ansatzes. Another approach to increase the expressibility of evolution is to go from a gate-based algorithm to a pulse-based algorithm, where the parameters $\theta$ take the form of physical control functions $z$, such as laser intensities or electrical currents \cite{Meitei2021,qocvqe2}. These have demonstrated the ability to realize a larger class of unitaries in less running time, mitigating errors, and increasing fidelity \cite{vqoc}. \\
 
\begin{figure}[b]
    \centering
    \includegraphics[width=0.90\linewidth]{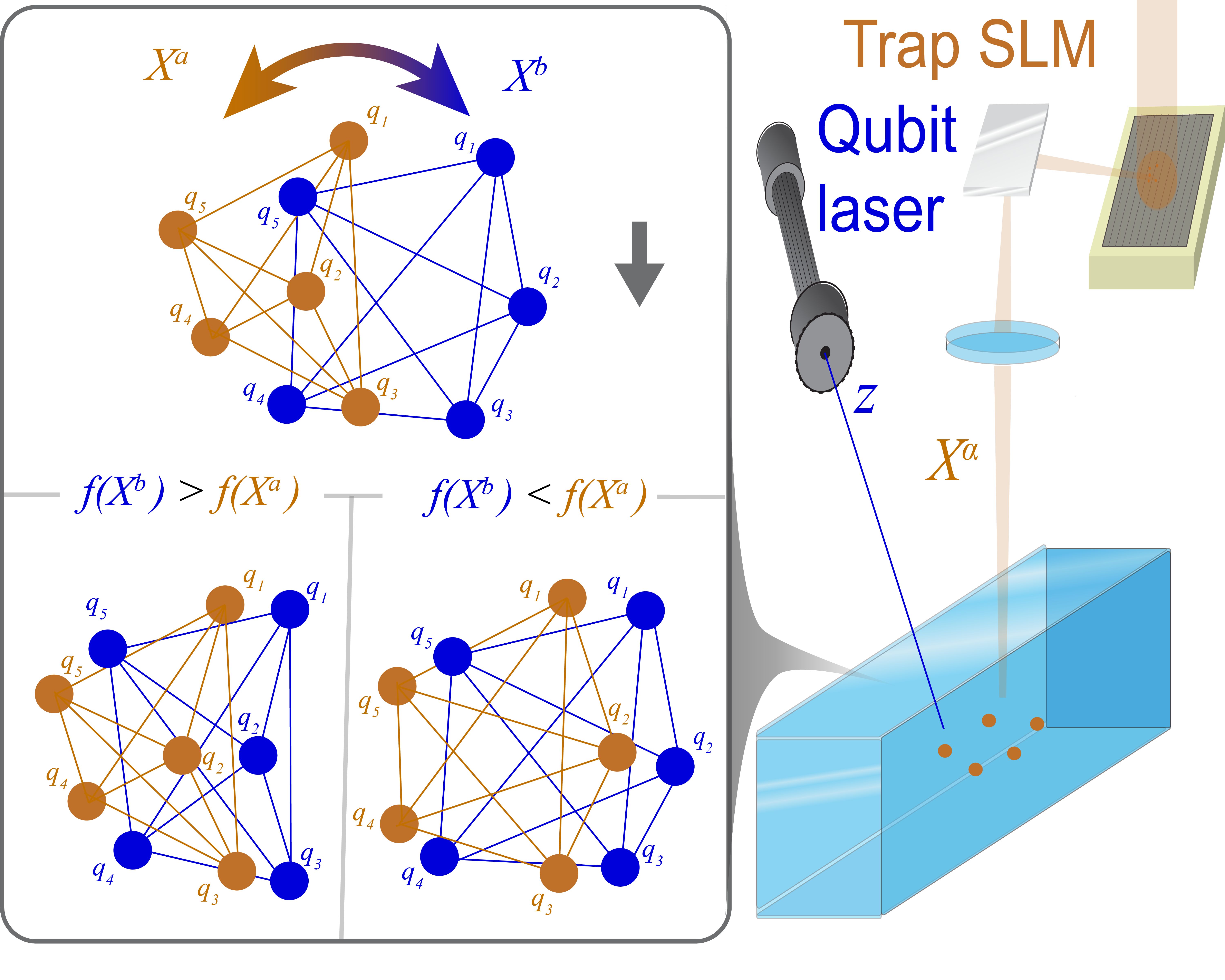}
    \caption{CBO of 5-qubit positions $X^\alpha$. On the right side, through the consensus based algorithm, the gold traps $X^ 
   a$ shift more to the blue traps $X^b$ than vice versa, since $f(X^b)<f(X^a)$. Left side shows analogous case for $f(X^b)>f(X^a)$. Positions of traps generated via SLM, and pulses $z$ executed through qubit lasers.}
    \label{fig:graphical abstract}
\end{figure}

 The choice of entangling operation in the ansatz can play an important role in the convergence rate of a given VQA \cite{entanglementchoice1,entanglementchoice2}. In addition, good initializations (including the qubit positions) have been shown to lead to barren plateau avoidance \cite{barrenplateau3,barrenplateau4,hatfunctions}, which are large flat regions of the parameter-cost function landscape that inhibit trainability \cite{barrenplateau1,barrenplateau2}. In this work, we leverage the unique ability of neutral atom tweezer platforms to realize arbitrary qubit position configurations. These positions determine the interactions between the qubits and subsequently the entanglement operation. This results in the possibility of tailoring qubit configurations to accelerate the convergence of pulse optimization. However, finding an optimal configuration for a particular problem Hamiltonian $H_{\text{targ}}$ is generally a difficult problem, since pulses are optimized only after the positions have been fixed.\\ 
 
 \textit{Relation to previous work -} Several works have previously considered optimizing qubit interactions for specific problems. In Ref.~\cite{topologyopt1}, a neural network is trained to select the problem-specific optimum from a finite number of possible configurations. As shown in Refs.~\cite{maxindependentset1, maxindependentset2}, graph problems such as maximum independent set inherently map well to the Rydberg Hamiltonians of neutral atom systems. Refs.~\cite{topologyopt2,topologyopt4} seek to optimize the atom positions with respect to the underlying structures of these graphs. In Ref.~\cite{topologyopt3}, an optimal qubit configuration is selected by optimizing a graph with weights related to the connectivity required by an input quantum circuit. Notably, none of these references use the gradient of the positions in their optimization schemes.\\ 
 
 This is logical, as the pulses are optimized with the underlying qubit positions in mind. Therefore, the gradients for the positions will be negligible. Furthermore, in neutral-atom systems the interaction strength scales with either $R^{-3}$ for dipole-dipole interactions or $R^{-6}$ for Van der Waals interactions \cite{morgado}, where $R$ is the distance between a pair of qubits. This leads to orders of magnitude difference between the gradient sizes of individual qubit pairs, see Sec.~\ref{sec:pulseopt}. As a result, the already small gradients are often focused on the interaction of one pair of qubits. Instead of a gradient-based approach, we opt to use a \textit{consensus-based optimization (CBO)} algorithm to optimize the qubit positions for specific target Hamiltonians $H_{\text{targ}}$ \cite{consensus}. Our algorithm initializes several `agents' $X^{(k)}$, which sample the parameter space of positions $\mathcal{X}$. Each of the agents partially optimizes the control pulses $z^{(k)}\in\mathcal{Z}$ with respect to their qubit positions $X^{(k)}$ to obtain an indication of the pulse-energy landscape. Through the consensus-based algorithm, this information is communicated across the agents to update the configurations for a subsequent iteration. After several iterations, the positions converge to a single configuration and the agents have reached a \textit{consensus}. We find that this optimized configuration generally gives a large improvement in the system's ability to solve the ground state minimization problem, as well as a significant acceleration in convergence.\\

The manuscript is structured as follows. Section~\ref{sec:pulseopt} provides an overview of the pulse optimization algorithm VQOC used in this work, as well as a similar gradient-based optimization for positions, highlighting its shortcomings. Section~\ref{sec:algorithm} presents our CBO methodology along with the numerical scheme used to solve for the optimal positions. Section~\ref{sec:results} shows the initial findings of our algorithm applied to random Hamiltonians and several small molecules. In Sec.~\ref{sec:conclusion}, we summarize our results and look at further research.

\section{Gradient based pulse optimization}
\label{sec:pulseopt}
The goal of the energy minimization problem is to prepare the ground state $|\psi_g\rangle$ such that $\langle\psi|H_{\text{targ}}|\psi\rangle\ge \langle \psi_g|H_\text{targ}|\psi_g\rangle=E_g$ for all states $\psi$, where $E_g$ is the ground state energy. The pulse optimization problem for a fixed configuration $X$ can be formulated as
\begin{equation}
    \min_{z\in \mathcal{Z}} J(X,z) := \langle \psi(T)|H_{\text{targ}}|\psi(T)\rangle+\mu\|z\|^2,
\end{equation}
where $|\psi\rangle=|\psi(X,z)\rangle$ satisfies the Schr\"odinger equation
\begin{equation}
\label{eq:hamiltoniaoc}
    \mathrm{i}\partial_t |\psi(t)\rangle = \big(H_V[X]+H_c[z]\big) |\psi(t)\rangle, \quad |\psi(0)\rangle=|\psi_0\rangle.
\end{equation}
Here, $H_V[X]$ is the interaction Hamiltonian, determined by the qubit positions $X\in\mathcal{X}$, and $H_c[z]$ is the control Hamiltonian, determined by the control functions $z\in\mathcal{Z}$. For $\mu>0$, the problem regularizes for the strength of the pulses $z$. This parameter can be raised to ensure that the maximum amplitudes of the found pulses lie within experimentally feasible ranges. We specify
\begin{equation}
\begin{aligned}
&\mathcal{Z}:=\bigg\{ z\in L^2((0,T);\mathbb{C}^L) \Big|\;\sup_{t\in[0,T]}|z_l(t)|\le z_{\max}\;\bigg\}, \\
    &\mathcal{X}:=\big\{\;[x^1,\dots,x^m]\; |\; x^i\in\mathbb{R}^2\big\}\simeq \mathbb{R}^{m\times 2},
\end{aligned}
\end{equation}
with $m$ the number of qubits. Note that since the atoms cannot be rearranged during the evolution of the state, $H_V[X]$ is time independent. Details on the control and interaction Hamiltonians for a neutral atom system can be found in App.~\ref{app:hamiltonians}.\\

As in Ref.~\cite{vqoc}, when $H_c[z(t)]=\sum_l^L z_l(t)H_l$ and $X$ is a fixed configuration, a gradient for this functional can be found as
\begin{equation}
\label{eq:derivativeVQOC}
\begin{aligned}
    &\nabla_z J(X,z)[\delta z_l] = -\mu \int_{0}^T z_l(t)\delta z_l(t)\di t \\
    &-2\mathrm{i}\int_{0}^T \big\langle \psi(t)\big| \big[H_j^\dagger,\, \Gamma^\dagger(T,t) H_{\text{targ}} \Gamma(T,t)\bigr]\big|\psi(t)\big\rangle\delta z_l(t)\, \di t,
\end{aligned}
\end{equation}
where $\delta z_l$ is a perturbation, $\Gamma(t,s):=U(t)U^\dagger(s)$, and $U$ is the unitary solution operator satisfying $|\psi(t)\rangle=U(t)|\psi_0\rangle$. This gradient can be used to iteratively optimize the pulses with respect to the cost function $J$ by taking steps in the direction of the gradient. We denote the pulses found after $n$ iterations as $z^n$.\\

Similarly, when $H_V[X]=\sum_{i\neq j} V_{ij}/\|x^i-x^j\|^p$ for $p\in\mathbb{N}$ (see App.~\ref{app:hamiltonians}), a gradient can be found for the positions of the qubits as
\begin{equation}
\begin{aligned}
    \nabla_X &J(X,z)[\delta X^l]=-2\mathrm{i}\sum_{i\neq j}\frac{p(x^i-x^j)}{\|x^i-x^j\|^{p+2}}\cdot \delta x^j\\
    &\times\int_{0}^T \big\langle \psi(t)\big|\big[V_{ji}^\dagger,\Gamma(T,t)^\dagger H_{\text{targ}}\Gamma(T,t)\big]\big|\psi(t)\big\rangle  \di t.
    \end{aligned}
\end{equation}
The problem with gradient based optimization becomes apparent from the fact that certain components will diverge when atoms come too close, especially for larger $p$. As a result, in many cases, only one pair of atoms will significantly contribute to the gradient. Furthermore, to find out whether under a configuration $X\in\mathcal{X}$ the ground state can be approximated, the pulses have to be optimized to some $X$-dependent optimal value $\tilde{z}_X$. Afterward, the gradients on the configuration $\nabla_X J(X,\tilde{z}_X)$ is very likely 0, as the pulses are optimized with the configuration $X$ in mind.\\


\section{Consensus-based algorithm}
\label{sec:algorithm}

To circumvent the challenges associated with gradient-based configuration optimization, we have chosen to implement a gradient-free approach based on the consensus-based optimization (CBO) algorithm introduced in Ref.~\cite{consensus}. Here, several agents $\{X^{(k)}\}_k=\{[x^{(k)1},\dots,x^{(k)m}]\}_k$ explore the configuration space $\mathcal{X}$ in a nested process. In the inner loop, pulses $z$ are partially optimized to assess the quality of their configurations. In the outer loop, their cost function values are inputs for some weighted average over the configurations, which is used to settle on the next set of configurations. 
Note that the qubits are distinguishable, and therefore ordering matters. The minimization problem is given as
\begin{equation}
    \min_{X \in \mathcal{X}}\min_{  z\in \mathcal{Z}} J(X,z)\quad\text{subject to \eqref{eq:hamiltoniaoc}}.
\end{equation}
$K$ agents $X^{(k)}\in\mathcal{X}$ are initialized and sample the configuration space to evaluate a cost function $f$ to be minimized. The agents then update their configurations according to an evolution equation given by
\begin{equation}
\label{eq:consensus1}
    \di X_\tau^{(k)}=-\lambda\big(X_\tau^{(k)}-v_f\big) \di \tau+\sqrt{2} \sigma|X_\tau^{(k)}-v_f|\, \di W_\tau^{(k)},
\end{equation}
where $\di W_\tau$ is a 2d white noise for the $x$ and $y$ directions, $\sigma>0$ is a diffusion coefficient, $\lambda>0$ is a drift coefficient, and $v_f$ is a weighted average given by
\begin{equation}
\label{eq:weightedmean}
    v_f=\frac{1}{\sum_k \omega_f^\alpha(X_\tau^{(k)})} \sum_k X_\tau^{(k)} \omega_f^\alpha(X_\tau^{(k)}),
\end{equation}
with $\omega_f^\alpha$ an exponential weight
\begin{equation}
\label{eq:consensus2}
\omega_f^\alpha(y)=\exp (-\alpha f(y)), \quad \alpha>0.
\end{equation}
Note here that $\tau$ is a non-physical timescale for the configuration space evolution and thus fundamentally different from the physical pulse timescale denoted by $t$. By invoking the Laplace principle from large deviation theory \cite{largedeviation}, it is possible to show that for $K\rightarrow \infty, \alpha\rightarrow \infty$ and $\sigma\rightarrow0$ we have that the distribution of the agents $X_\tau^{(k)} \sim \rho_\tau\rightarrow \delta({\arg \min}_{x\in\mathcal{X}} f(x))$ in a distributional sense \cite{laplace2}. And thus all agents will reach a consensus exactly on the global minimum. By adding the noise $\sigma$-term, local minima can be avoided, which has heuristically shown to improve convergence in many cases \cite{Neelakantan2015AddingGN}. In Ref.~\cite{consensus}, the drift $\lambda$-term is multiplied by a regularization of the Heaviside function on $f(X^{(k)}_\tau)-f(v_f)$ in order to guarantee functional descent. We have found this term does not significant change our results, and for simplicity have left it out.\\

The discretized optimization procedure is given by
\begin{equation}
    X_{n+1}^{(k)}=X_{n}^{(k)}-\lambda(X_{n}^{(k)}-v_f) \Delta \tau+\sqrt{2} \sigma|X_n^{(k)}-v_f| \mathcal{N}\sqrt{\Delta \tau},
\end{equation}

\begin{figure}[b]
    \centering
    \includegraphics[width=0.85\linewidth]{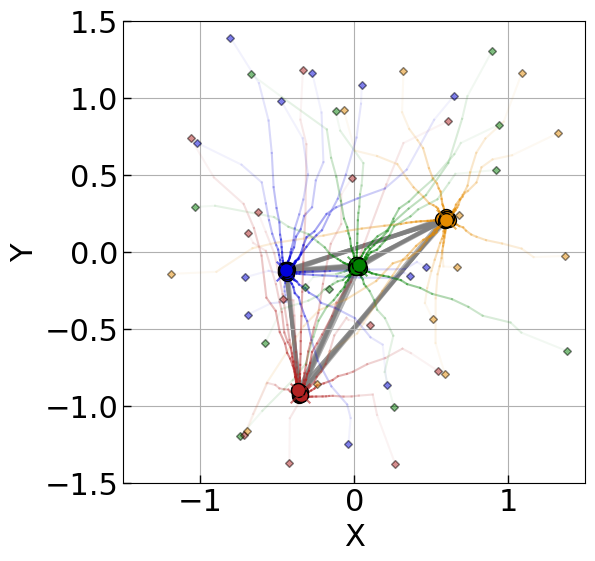}
    \caption{Example of configuration evolution for a 4-qubit problem with 12 agents, by means of CBO. Agents are initialized at $\tau=0$, communicate information on the minimal energy achieved. Configurations are updated until consensus is achieved at $\tau=1$. Qubits are colorized by index.}
    \label{fig:positionoptimizaiton}
\end{figure}

where $\Delta\tau>0$, $\mathcal{N}$ is a standard normal distribution, and we finish after $N_{\text{out}}$ outer iterations. Ideally, we would choose $f=\min_{z\in\mathcal{Z}} \langle\psi(T)|H_{\text{targ}}|\psi(T)\rangle$. However, finding the global minimum for $z$ is strenuous and could take many iterations of VQOC. Because we are looking for an indication of how well a configuration can solve for the ground state energy $E_g$, we take $f=-\log(J(X,z^{N_{\text{in}}})-E_g)$ to be the log-energy error reached after a small number $N_{\text{in}}$ of VQOC iterations, so-called inner iterations. Choosing $N_{in}$ relatively small ensures that configurations that quickly decrease in energy are preferred. The $\log$ leads to a wider range in $f$ values for all individual agents, making the choice of $\alpha$ in the exponential weighting less stringent. The same behavior can in principle be achieved by leaving $f=J(X,z^{N_{\text{in}}})$ and taking $\alpha$ a higher value. This case is useful when (a good approximation) of $E_g$ is not known.\\

Alternatively, we could choose 
\begin{equation}
\label{eq:altcost}
    f=J(X,z^{N_{\text{in}}})-\nu \left\|\nabla_z J(X,z^{N_{\text{in}}})\right\|,
\end{equation}
for some regularizer $\nu>0$. Here, we actively optimize for the pulse gradient of $J$ to be large so that further optimization of the pulses will yield better results. The added benefit is that the algorithm prioritizes large-valued gradients, avoiding configurations with large regions of flat pulse-energy landscape, so-called barren plateaus \cite{barrenplateau1,barrenplateau2}. If there is minimal variation in the $f$-values among the agents, potentially due to the occurrence of barren plateaus, the consensus-based algorithm may struggle to identify advantageous new configurations, leading to stagnation in the algorithm’s progress for small diffusion coefficient $\sigma>0$. \\

\begin{algorithm}[h]
\SetAlgoLined
\SetInd{0.5em}{0.5em}
\SetKwData{Left}{left}\SetKwData{This}{this}\SetKwData{Up}{up}
\SetKwFunction{Union}{Union}\SetKwFunction{FindCompress}{FindCompress}
\SetKwInOut{Input}{input}\SetKwInOut{Output}{output}
\Input{$z^{(k)}_{0,0}$, $X^{(k)}_0$, $H_{\text{targ}}$, $\phi_0$, $N_\text{out}$,$N_{\text{agents}}$, $N_\text{in}$, $N_\text{final}$}
\Output{$X_{N_\text{out}}^{(k)},z^{(k)}_{N_\text{out},N_{\text{final}}}, J(X_{N_\text{out}}^{(k)},z^{(k)}_{N_\text{out},N_{\text{final}}})$}
\BlankLine
\tcp{Configuration Optimization}
\tcp{Outer Loop}
\For{$n=0$ \KwTo $N_{\text{out}}$}{
\For{$k=0$ \KwTo $ N_\text{agents}$}{
\tcp{Inner Loop}
\For{$l=0$ \KwTo $N_{\text{in}}$}{
$z^{(k)}_{n,l+1}=z^{(k)}_{n,l}-\gamma\nabla_z J(X_n^{(k)},z^{(k)}_{n,l})$;
}
calculate $J(X_n^{(k)},z^{(k)}_{n,N_\text{in}});$
}
calculate $v_f$ according to \eqref{eq:weightedmean};\\
draw normal random variables $\mathcal{N}^{n,k}$;\\    
$X_{n+1}^{(k)}=X_{n}^{(k)}-\lambda(X_{n}^{(k)}-v_f) \Delta \tau$ 
$\textcolor{white}{.}\quad\quad\quad\quad+\sqrt{2} \sigma|X_n^{(k)}-v_f| \mathcal{N}^{n,k}\sqrt{\Delta \tau}$;
}
\tcp{Final Energy Determination}
\For{$k=0$ \KwTo $ N_\text{agents}$}{
\For{$l=0$ \KwTo $N_{\text{final}}$}{
$z^{(k)}_{N_\text{out},l+1}=z^{(k)}_{N_\text{out},l}-\gamma\nabla_z J(X_{N_\text{out}}^{(k)},z^{(k)}_{N_\text{out},l})$;
}
calculate $J(X_{N_\text{out}}^{(k)},z^{(k)}_{N_\text{out},N_{\text{final}}})$;
}
return $X_{N_\text{out}}^{(k)},z^{(k)}_{N_\text{out},N_{\text{final}}}, J(X_{N_\text{out}}^{(k)},z^{(k)}_{N_\text{out},N_{\text{final}}})$;\\
\caption{Configuration optimization}\label{algo:algo1}
\end{algorithm}

Lastly, we comment on the scalability of the method. For pulse optimization, it is known that significantly more inner iterations are required when scaling up the number of qubits, mainly due to the larger dimensional control space \cite{vqoc}. For our method, we find that the number of outer iterations does not need to be increased with the number of qubits, as the configuration is encoded per agent rather than per qubit. It is likely that for an increasing number of qubits more agents are necessary as the dimensionality of the configuration space increases \cite{consensus2}. Because of the classical computational intensity of testing for a large number of qubits, this is outside the scope of this work. Lastly, our CBO algorithm offers a large advantage in terms of parallelization on current era neutral atom computers for two reasons. First, in a neutral atom system both the atom preparation and measurement take significantly more time compared to the evolution of the qubits \cite{madjarov}. Second, despite the fact that many atoms can be prepared on one chip, simultaneous control of these atoms is often limited to a few \cite{Wintersperger2023}. In our algorithm, all agents can be initialized and measured together, but evolved individually. This parallelizes the time-consuming preparation and measurement, while also satisfying the need to control only a few atoms simultaneously.

\section{Results}
\label{sec:results}

In this section, we present several examples to illustrate the performance of our configuration optimization using the consensus algorithm and provide comparisons with suitable random configuration counterparts. Unless stated otherwise, the hyperparameters for position optimization are taken as $(\alpha,\lambda,\sigma,\Delta \tau)=(4,0.4,0.1,0.5)$. These have empirically shown to lead to well-optimized configurations over a large scale of problems (see Sec.~\ref{sec:hyperparameter}). The generation of the initial configurations is specified in App.~\ref{app:positioninit}, after which $N_{\text{out}}=20$ outer iterations of the CBO algorithm are performed to produce the final configurations $X^{(N_\text{out})}$. Pulses $z_l$ are encoded as step functions with 100 equidistant steps between $t=0$ and $t=1$. In order to assess our method on many different problems, we sample random target Hamiltonians $H_{\text{targ}}=\sum_{i}\alpha_i P_i$ with $\alpha_i\sim \text{Unif}[0,1]$ and $P_i$ respectively random coefficients and random Pauli strings. Here, each Pauli string has a 20\% chance of being selected. By picking the coefficients and strings in this way, ground energies are found around a magnitude of $10^1$, resulting in $H_{\text{targ}}$ closely resembling realistic molecular Hamiltonians as expressed in atomic units \cite{kandala}. We will often consider a problem solved, once an energy error of $10^{-3}$ has been achieved, corresponding to the chemical accuracy measured in Hartree \cite{chemaccuracy}. For all simulations, we will take the interaction strength coefficient $C_{3,6}=1$ in arbitrary units (see App.~\ref{app:hamiltonians}). All subsequent pulse strengths and interatomic distances are expressed in terms of this interaction energy coefficient. Note that for other units, time can always be rescaled by the Schr\"{o}dinger equation \eqref{eq:hamiltoniaoc} so that $C_{3,6}=1$.\\ \\  

Figure~\ref{fig:positionoptimizaiton} shows an example of the configuration evolution for the CBO algorithm for 12 agents, 4 qubits and a randomly sampled $H_{\text{targ}}$. The initial configurations are sampled, after which they start to concentrate and eventually reach a consensus. All results in this work will initialize 12 agents, which is equal to the number of available cores on our classical simulation system, but the algorithm would obviously benefit from more agents.

\subsection{GHZ state preparation}
\begin{figure}[H]
    \centering
    \includegraphics[width=0.825\linewidth]{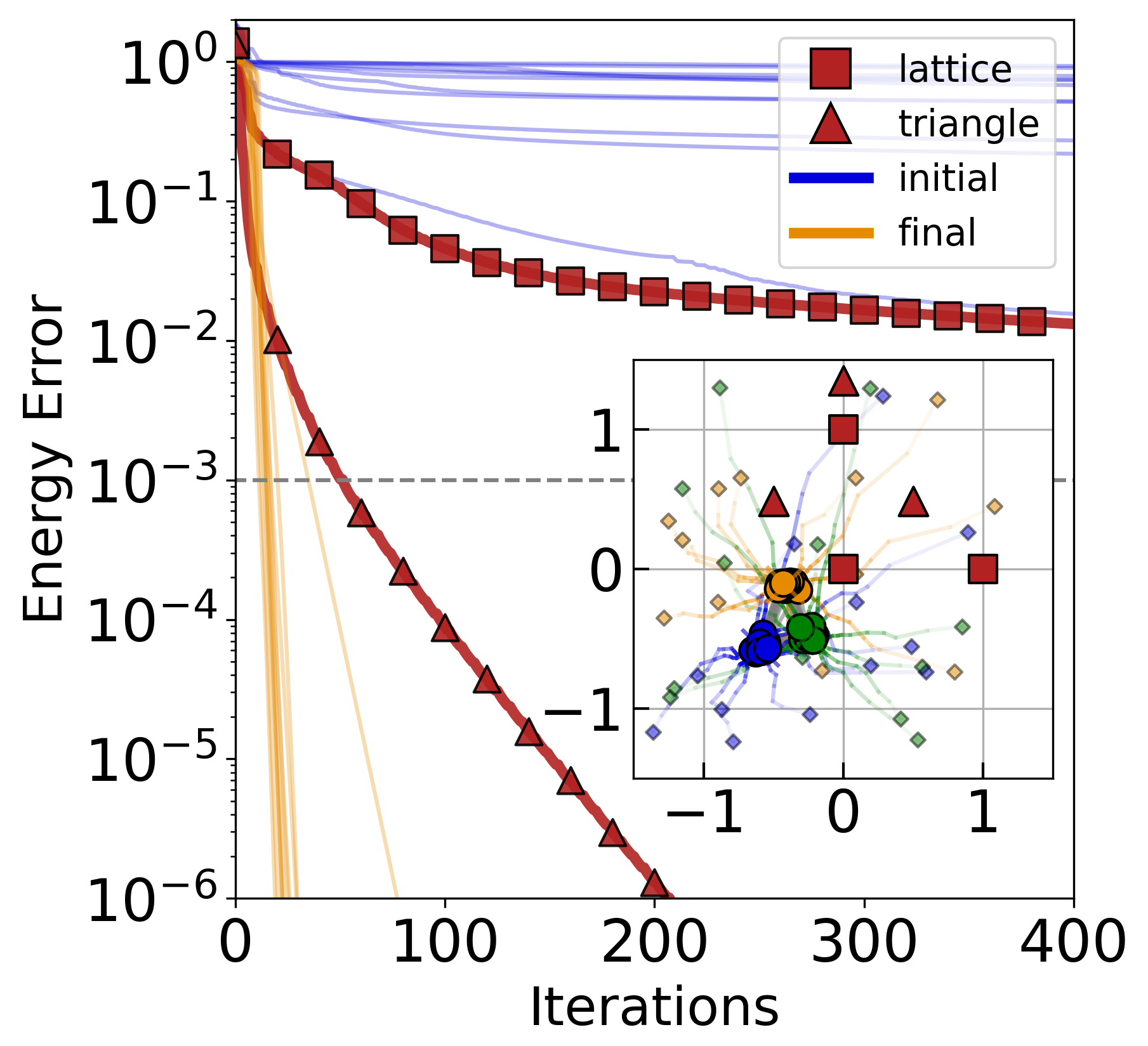}
    \caption{Energy error convergence for preparation of 3-qubit GHZ state for initial lattice (squares) and triangle (triangles) configurations (red). Random initial configurations (blue) are optimized (gold) and reaching much faster energy error convergence. Inset: lattice (red squares) and triangle (red triangles) configurations with initial to optimized configuration evolution. $N_{\text{in}}=100, N_{\text{out}}=20$. Colorized by qubit index.}
    \label{fig:ghz}
\end{figure}

As a first illustrative example, we investigate the preparation of a maximally entangled Greenberger–Horne–Zeilinger (GHZ) \cite{ghz} state on 3 qubits 
\begin{equation}
|\text{GHZ}\rangle=\frac{1}{\sqrt{2}}(|000\rangle+|111\rangle),
\end{equation}
by taking $H_{\text{targ}}=-|\text{GHZ}\rangle\langle \text{GHZ}|$. To show the resulting energies reached more clearly, we will solve the pulse optimization for $N_\text{final}\gg N_{\text{in}}$ iterations throughout this work for both the initial and final configurations. Figure~\ref{fig:ghz} shows that the pulse optimization for an equilateral triangle configuration (red triangles) performs much better than in a lattice configuration (red squares), which is logical given the symmetry in $H_\text{targ}$. This indicates that certain configurations exhibit significantly better performance compared to others. The CBO algorithm can be seen to take symmetry into account as it converges to an equilateral triangle-like configuration. The optimized interatomic distance in the equilateral triangle also contributes to reaching convergence.

\subsection{Interaction types}
To investigate the performance of our algorithm for varying interaction types, we run our consensus-based position optimization procedure for Dipole-Dipole, VdW gr-qubit, and VdW gg-qubit interactions (see App.~\ref{app:hamiltonians}). Figure~\ref{fig:interactions} shows representative instances for each of these interaction types. For both Dipole-Dipole (Fig.~\ref{fig:interactions}a) and VdW gg-qubit interactions (Fig.~\ref{fig:interactions}c), we see a great decrease in energy error for the optimized positions (gold) compared to the initial positions (blue). In particular, pulse optimization is somewhat slower for VdW gg-qubits due to the extra controls and the qutrit manifold $\{|0\rangle, |1\rangle, |r\rangle\}$ (rather than qubit). Figure~\ref{fig:interactions}b shows a typical instant for VdW gr-qubits. When inside each other's Rydberg blockade radius, two interacting qubits can not be excited to their $|1\rangle$ states simultaneously, and therefore the ground state can not be reached for most $H_\text{targ}$. Nevertheless, this use case clearly illustrates another advantage of the position optimization, which is that convergence is sped up. From Fig.~\ref{fig:interactions}b, we clearly see that fewer pulse optimization iterations are needed to reach the lowest possible energy. The advantage of the CBO is thus two-fold: configurations are found that lead both to lower energies and fewer necesary pulse iterations. For the rest of this section we will use Dipole-Dipole interacting qubits as these do not inhibit any part of the computational state space from being reached and thus give the clearest result on energy error improvement.

\subsection{Hyperparameter analysis}
\label{sec:hyperparameter}
Next, the influence of the hyperparameters $\alpha,\sigma$ and $\lambda$ on the consensus algorithm is investigated, as in \eqref{eq:consensus1} and \eqref{eq:consensus2}. Figure~\ref{fig:hyperparameters} shows the results for varying hyperparameters, but with the exact same initial configurations and target Hamiltonians $H_{\text{targ}}$. For all cases, solutions are found below chemical accuracy. However, the behavior of the solutions differs strongly. In Fig.~\ref{fig:hyperparameters}c, the parameter $\alpha$ is enhanced, leading to higher weights $\omega_f^\alpha$ for better solutions. This results in a stronger and faster convergence and less overall exploration of the configuration space. In Fig.~\ref{fig:hyperparameters}d, $\sigma$ is enhanced, leading to more diffusion and total exploration of the configurations at the cost of slower consensus. This can be seen in Fig.~\ref{fig:hyperparameters}a, where the final errors still vary quite strongly after 100 pulse optimizations. Lastly, in Fig.~\ref{fig:hyperparameters}e, $\lambda$ is enhanced, giving rise to faster attraction to the weighted mean. For this case, we see in Fig.~\ref{fig:hyperparameters}a that all configurations have very similar pulse optimizations that come at the cost of exploration. In general, we empirically find that the hyperparameters $(\alpha=4,\sigma=0.1,\lambda=0.4)$ lead to good solutions. The final configurations found for all of these different cases are similar in configuration, with the exception of Fig.~\ref{fig:hyperparameters}b, indicating the existance of several well-performing configurations for a particular instance of $H_{\text{targ}}.$

\begin{widetext2}
    \begin{minipage}[b]{\linewidth}
        \begin{figure}[H]
            \centering
            \includegraphics[scale=0.47]{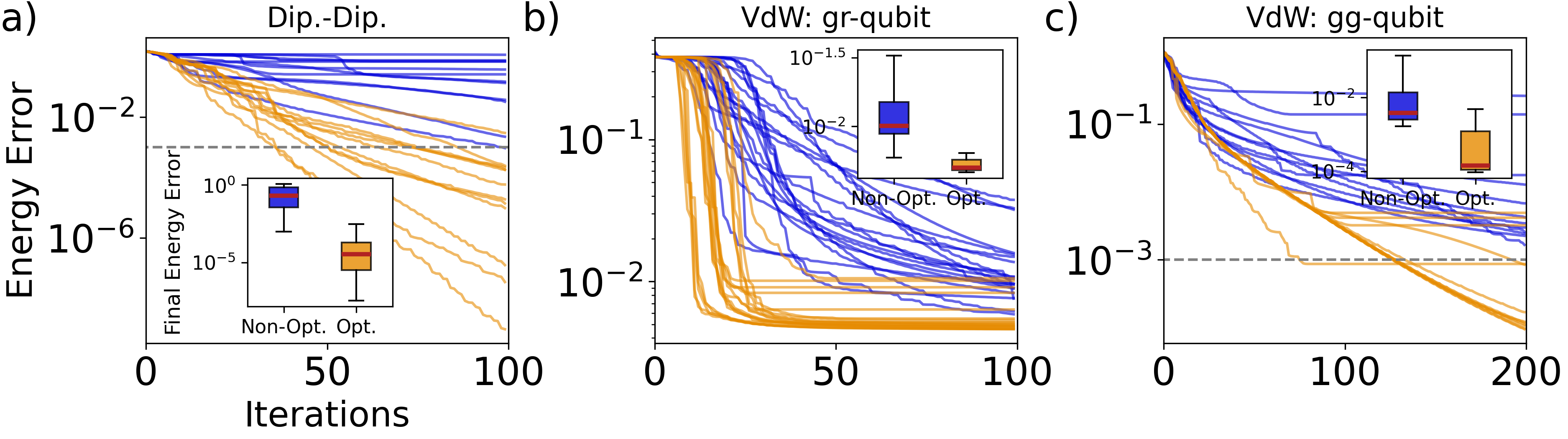}
    \caption{Examples of CBO for a randomized target Hamiltonian under different types of interactions. Shown is the pulse optimization energy errors for 12 agents with non-optimized (blue) and optimized positions (gold). Energy errors stay high for the non-optimized positions, whereas several agents for the optimized positions reach low errors. a) Dipole-Dipole energy interactions, showing order magnitude lower errors. b) VdW gr-interaction, illustrating faster convergence when full optimization is not possible. c) VdW gg-interaction, showing low errors for qutrit system. For all cases, $N_{\text{in}}=20, N_{\text{out}}=20$.}
            \label{fig:interactions}
        \end{figure}    
    \end{minipage}
\end{widetext2}

\begin{widetext2}
    \begin{minipage}[b]{\linewidth}
        \begin{figure}[H]
            \centering
            \includegraphics[scale=0.520]{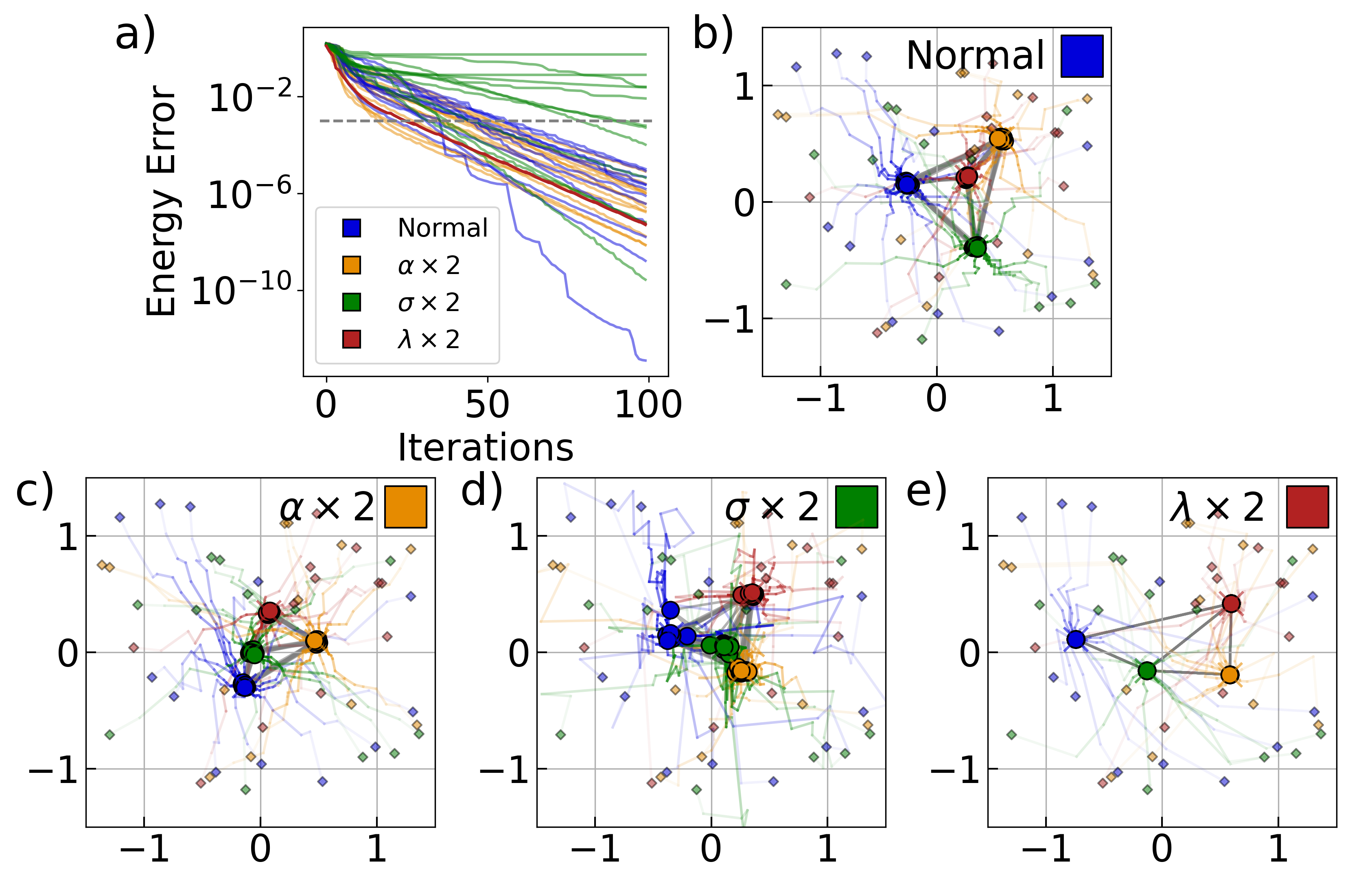}
    \caption{Varying of hyperparameters in CBO algorithm. a) pulse energy optimization for varying hyperparameters. b-e) position evolution of varying hyperparameters, with color of top right square indicating corresponding pulse evolution in a). b)  normal hyperparameters of $\alpha=4,\sigma=0.1,\lambda=0.4$. c-e) position evolution for respective adjustment of hyperparameters $\alpha, \sigma$ and $\lambda$. Qubits are colorized by index. For all cases, $N_\text{in}=20, N_\text{out}=100.$}
            \label{fig:hyperparameters}
        \end{figure}    
    \end{minipage}
\end{widetext2}

\subsection{Randomized Hamiltonians}

In this section, we illustrate the consistency of our method by solving for a fixed $H_{\text{targ}}$ for 20 instances of initial configurations. We then repeat this for several $H_{\text{targ}}$, each of which constitutes an `individual problem'. We want to compare the final results with the randomized initial configurations. However, as seen from Figs.~\ref{fig:positionoptimizaiton} and \ref{fig:hyperparameters}, the final configuration must always lie in the simplex of the originally initialized configurations (unless $\sigma\gg \lambda$). Therefore, the final configuration will on average have smaller interatomic distances and thus create more entanglement. Thus, it is not entirely fair to compare the random initializations to the final contracted configuration.\\

For this reason, random configurations that are more comparable to the final configurations are desired. Since all initial configurations and $H_{\text{targ}}$ considered are drawn from the same distributions, we will estimate a probability density function of all interatomic distances between the final configurations.

\begin{figure}[t]
    \centering
    \includegraphics[width=0.95\linewidth]{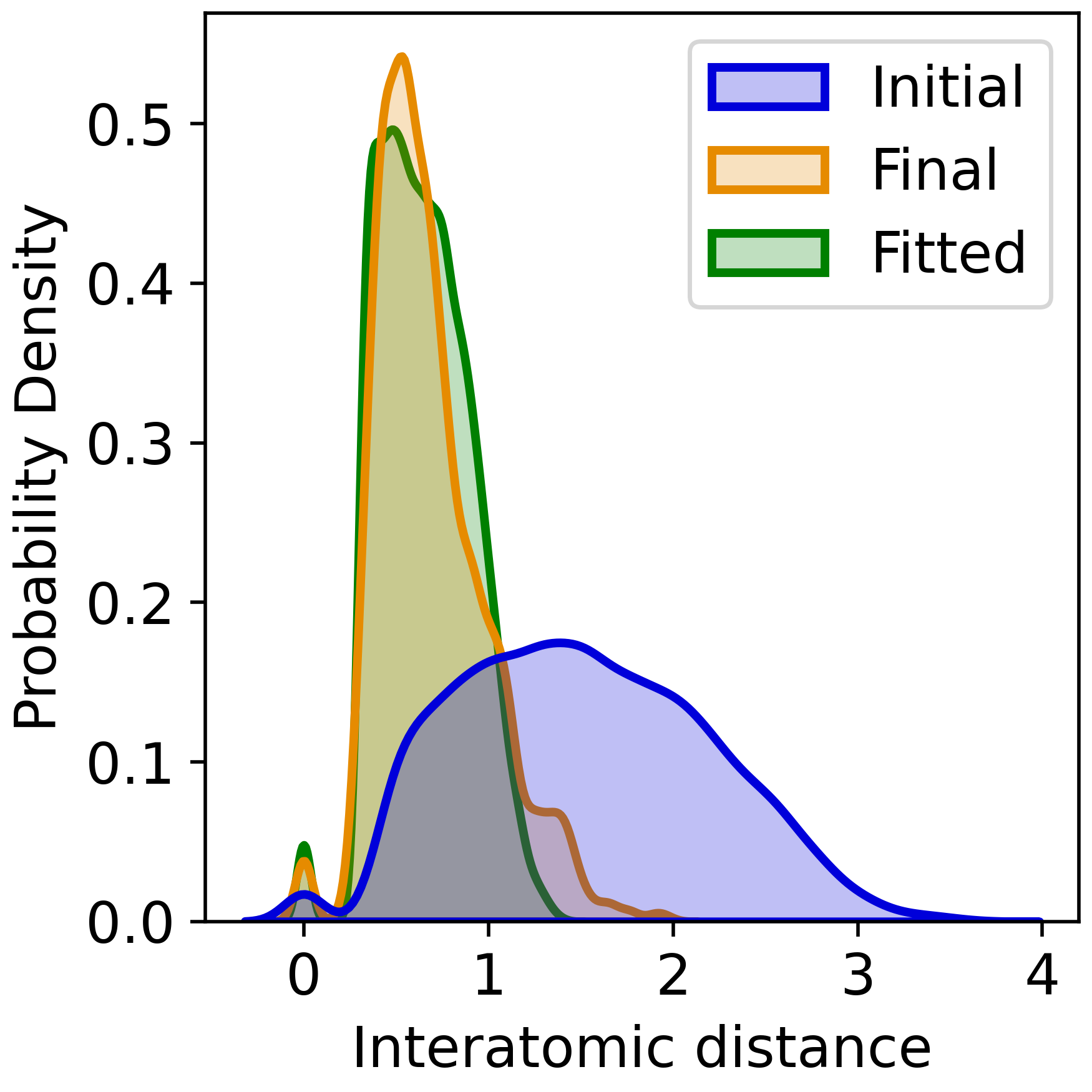}
    \caption{Probability density functions for the pairwise interatomic distances. The pairwise distances for initial configurations (blue, as described in App.~\ref{app:positioninit}), for final configurations after running the algorithm (gold) and for configurations fitted to the final configurations (green, as described in App.~\ref{app:positioninit}).}
    \label{fig:distances}
\end{figure}

A method of creating initial configurations can then be fitted so that it generates an equivalent distribution, as described in App.~\ref{app:positioninit}. This leads to random configurations that will have interatomic distance distributions similar to the final configurations and thus are more fair to compare against. We will call these the \textit{fitted} configurations, see Fig.~\ref{fig:distances}.\\

Figure~\ref{fig:multiple} shows the statistics on the multiple random target Hamiltonians. From Fig.~\ref{fig:multiple}a, we see that there is a large decrease in error between the final and fitted configurations. Figure~\ref{fig:multiple}b shows most clearly that over many different initializations and many different $H_{\text{targ}}$, our method can consistently find configurations leading to lower energy errors. In many cases the initial configurations do not reach chemical accuracy. However, for the fitted and the final configurations, chemical accuracy is reached in almost all cases, with the final configurations often still outperforming the fitted ones by several orders of magnitudes. 

\subsection{Molecular Hamiltonians}
\label{sec:molecules}

Lastly, our method is tested for practical applications by minimizing the energies of small molecules, where their internal structure is varied. For this, molecular Hamiltonians are generated using the Psi4 quantum chemistry library \cite{psi4}. We consider LiH with varying distances between Li and H, resulting in 4-qubit Hamiltonians. CH$_4$ in the 2D plane with varying distances between the C and H atoms, resulting in 5-qubit Hamiltonians. Lastly, BeH$_2$ for varying distances between Be and the two H's resulting in a 6-qubit Hamiltonian.\\

The results are shown in Fig.~\ref{fig:molecules}. For all three molecules, configurations are found that largely outperform their fitted configuration counterparts. It was found that in order to solve the problems, the number of inner iterations needed to be scaled with the number of qubits, which has been reported before for pulse optimization algorithms and VQAs in general \cite{vqoc,iterations1,iterations2}. However, as mentioned in Sec.~\ref{sec:algorithm}, the position optimization does not suffer from this, and the number of outer iterations can be kept at $N_{\text{out}}=20$. The CH$_4$ and BeH$_2$ ground states also seem easier to find than the LiH ground state. A reason for this could be the fact that LiH has a more entangled ground state than the other two molecules \cite{vqoc}. In future work, it would be interesting to quantify how well a configuration for one fixed interatomic distance functions for another one close to it. \\

\begin{widetext2}
    \begin{minipage}[b]{\linewidth}
        \begin{figure}[H]
            \centering
            \includegraphics[scale=0.35]{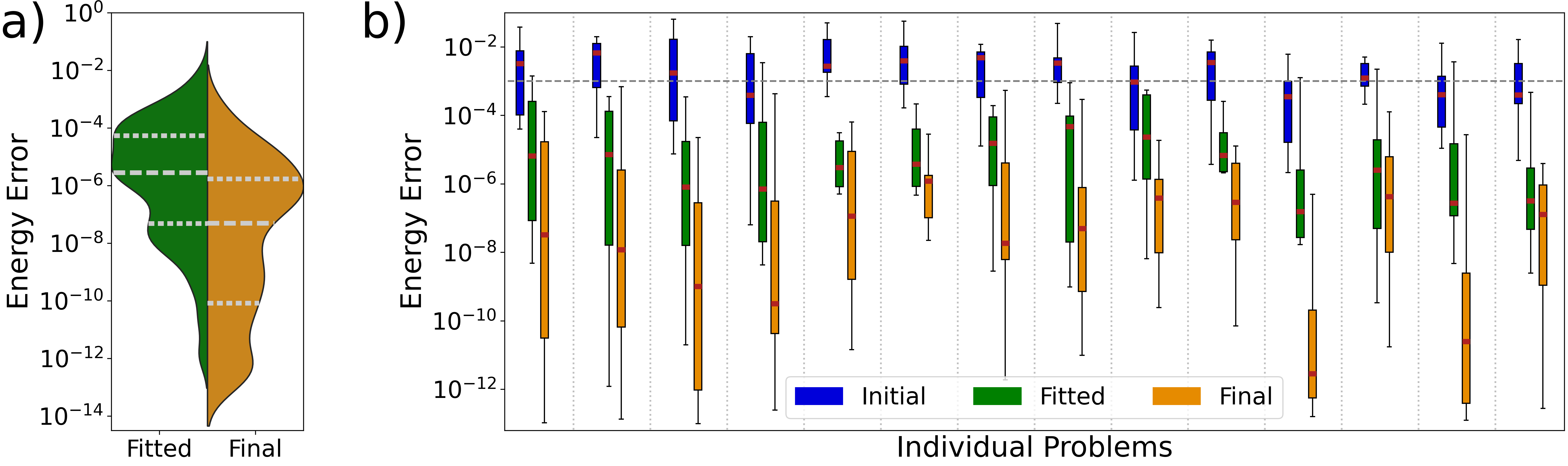}
    \caption{CBO results for 14 random target Hamiltonians, with 4 qubits, 20 agents and dipole-dipole interactions, repeated 20 times each. $N_{\text{out}}=20$ outer iterations and $N_{\text{in}}=100$ inner iterations. a) distribution of log errors for final configurations (gold) and fitted configurations (green). b) Log errors for initial (blue), fitted (green) and final (gold) configurations, separated for all 14 individual problem.}
            \label{fig:multiple}
        \end{figure}    
    \end{minipage}
\end{widetext2}

\begin{widetext2}
    \begin{minipage}[b]{\linewidth}
        \begin{figure}[H]
            \centering
            \includegraphics[scale=0.48]{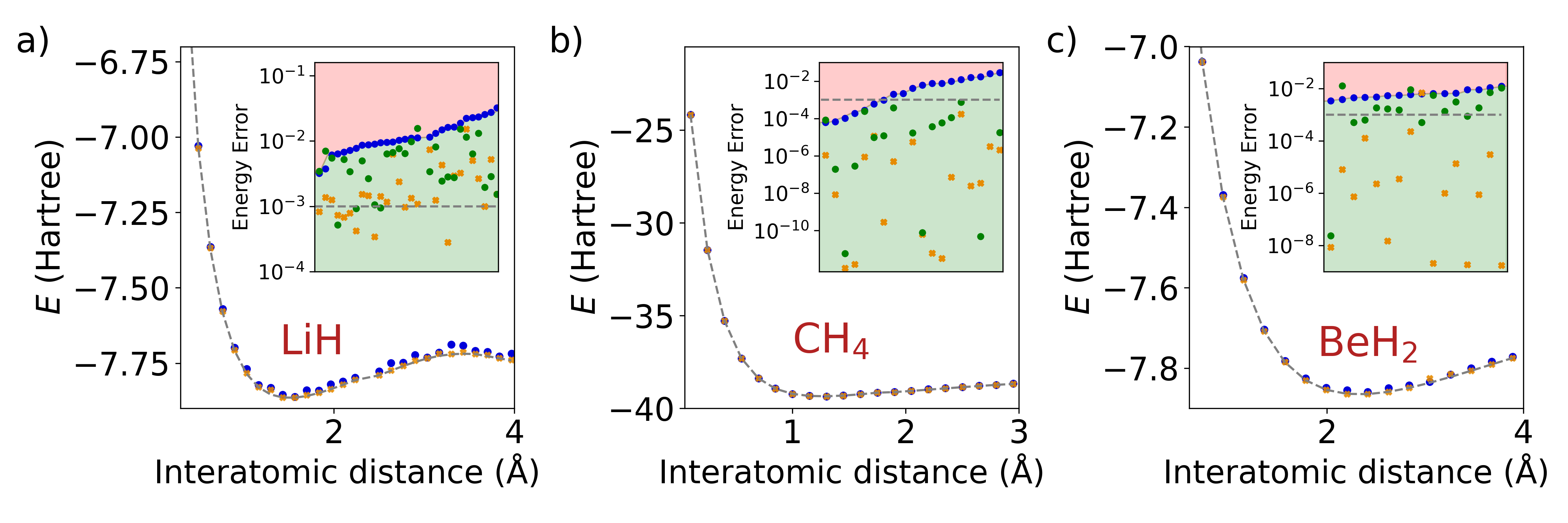}
    \caption{CBO results for several small molecules. For all molecules, we have 20 agents and $N_{\text{out}}=20$ outer iterations, shown are non-optimized positions initialized far apart (blue), optimized positions from blue initialization (gold), and non-optimized positions initialized according to distance distribution of gold optimized positions (green). Results in insets sorted on $x$-axis based on blue results for visibility. a) LiH (4 qubits, $N_\text{in}=50$), b) CH$_4$ (5 qubits, $N_\text{in}=100$), c) BeH$_2$ (6 qubits, $N_\text{in}=200$).  }
            \label{fig:molecules}
        \end{figure}    
    \end{minipage}
\end{widetext2}

\section{Conclusion}
\label{sec:conclusion}
This work discusses and analyzes a new method to construct problem-adapted configurations for variational quantum algorithms. This is an important issue, as the choice of entanglement ansatz, determined by the position-dependent interaction strength between qubits, plays an important role in both the solvability of the problem and the avoidance of barren plateaus. In particular, we leverage the unique advantage of a neutral atom tweezer platform to be able to place qubits anywhere in the 2D plane, thus having the ability to create arbitrary configurations. Gradient optimization of the qubit positions has shown to be hard as a consequence of the divergent nature of the interaction energies. Instead, we opt to use a gradient-free CBO algorithm to find improved qubit configurations. Our work shows that this consensus-based algorithm is able to effectively sample the configuration space to find qubit configurations that lead to faster and lower error solutions of the ground state minimization problem, for a large scale of random and practical Hamiltonians. We also hypothesize that our algorithm can be used practically unchanged for other quantum optimization algorithms.\\

In future work, we hope to improve the cost functions used in the CBO algorithm to use the size of the gradient as information, as in Eq.~\eqref{eq:altcost}. For our found solutions, we generally see steeper gradients and faster convergence in pulse optimization. In view of this, our aim is to quantitatively investigate the effect our optimized configurations have on the avoidance of barren plateaus. Another interesting problem would be to examine the correlations between optimized configurations and the target Hamiltonian $H_{\text{targ}}$. This could reveal information on better configuration ansatzes. Lastly, experimental verification of the optimized configurations versus randomized configurations could be of great interest. 

\section*{ACKNOWLEDGEMENTS}
We thank Bas Wijnsma, Raul dos Santos, Emre Akaturk, Jasper Postema, Madhav Mohan and Jasper van de Kraats for fruitful discussions. This research is financially supported by the Dutch Ministry of Economic Affairs and Climate Policy (EZK), as part of the Quantum Delta NL program, the Horizon Europe programme HORIZON-CL4-2021-DIGITAL-EMERGING-01-30 via the project 101070144 (EuRyQa), and by the Netherlands Organisation for Scientific Research (NWO) under Grant No.\ 680.92.18.05 and NGF.1582.22.009.

\section*{COMPETING INTERESTS}
The authors declare no competing interests.\\

\section*{DATA AVAILABILITY}
The data supporting the findings are available from the corresponding author upon reasonable request.

\section*{CODE AVAILABILITY}
The code supporting the findings is available from the corresponding author upon reasonable request.

\newpage
\bibliographystyle{apsrev4-1}
\bibliography{Bibliography.bib}
\newpage
\onecolumngrid

\appendix

\section{Position Initialization}
\label{app:positioninit}
Here, we describe the generation of initial configurations. As the tweezer positions are given in the 2D plane, a configuration $X$ for $m$ qubits is given by $X=[x^{1},\ldots x^m]\in\mathbb{R}^{m\times2}.$ The qubits are placed randomly within a box $[-R_{\text{max}},R_{\text{max}}]^{2}$, where the qubit positions get resampled whenever the euclidean distance between any two qubits is below $R_{\text{min}}>0$. The first atom can thus be placed anywhere in $[-R_{\text{max}},R_{\text{max}}]^{2}$ and each subsequent atom will be repeatedly placed at random in $[-R_{\text{max}},R_{\text{max}}]^{2}$ until it is at least distance $R_{\text{min}}$ from the already placed atoms.\\

As mentioned in Sec.~\ref{sec:results}, configurations tend to contract under the consensus algorithm. This leads to lower interatomic distances and thus facilitates more entanglement. Therefore, it is not entirely fair to compare initial configurations to the final ones. By only varying the target Hamiltonian $H_{\text{targ}}$, we get a large collection of similar problems. We can find the kernel density estimate of interatomic distances to get the gold curve in Fig.~\ref{fig:distances}. We can then optimize for a new $R_{\text{min}}$ and $R_{\text{max}}$ to get a distribution of interatomic distances (Fig.~\ref{fig:distances} green) that most resembles that of the final configurations (Fig.~\ref{fig:distances} gold). This is done by minimizing the Kullback-Leibler divergence \cite{kullback} between the two distributions.

\section{Rydberg neutral atoms}
\label{app:hamiltonians}

This section introduces basic Rydberg physics to identify what control pulses and especially interactions can look like for this system, as discussed in Sec.~\ref{sec:results}. This will yield both the control Hamiltonian $H_c[z]$ depending on the pulses, and the interaction Hamiltonian $H_V[X]$ depending on the configuration, see \eqref{eq:hamiltoniaoc}. We consider a neutral atom quantum computing platform consisting of individual neutral atoms trapped in optical tweezers, where the electronic states encode for the qubit manifold \cite{rydberg1}. Generally, three states are considered for neutral atom systems, a well-isolated manifold consisting of the ground state $|g_0\rangle$ and a meta-stable state $|g_1\rangle$, as well as an auxiliary Rydberg state $|r\rangle$ used for interaction.

\medskip

Single qubit rotations on qubit $j$ between states $|a\rangle$ and $|b\rangle$ are facilitated by a laser interacting with the atom to realize the Hamiltonian \cite{morgado,rydberg1}
\begin{equation}
\begin{aligned}
    H_{j}^{ab}=\frac{\Omega_{ab,j}(t)}{2} \left(e^{\mathrm{i} \varphi_{ab,j}(t)}|a\rangle_j\langle b|_{j}+e^{-\mathrm{i} \varphi_{ab,j}(t)}|b\rangle_j\langle a|_{j}\right)-\Delta_{b,j}(t)|b\rangle_{j}\langle b|_j.
\end{aligned}
    \label{eq:qubitlightinteraction}
\end{equation}
On atom $j$, $\Omega_{ab,j}(t)$ denotes the coupling strength, $\varphi_{ab,j}(t)$ the phase of the coupled laser, and $\Delta_{b,j}(t)$ = $\omega_{ab,j}(t)-\tilde{\omega}_{ab}$ the detuning of the laser frequency $\omega_{ab,j}(t)$ from the energy level difference $\tilde{\omega}_{ab}$. In current Rydberg systems, one has less control over $\varphi$ than over $\Omega$ and $\Delta$, \cite{phase1}, and subsequently we set $\varphi=0$. For our systems, we assume transitions $|g_0\rangle\leftrightarrow|g_1\rangle$ and $|g_1\rangle\leftrightarrow|r\rangle.$ This renders control pulses $z(t)\in\{\Omega_{g_{0}g_{1},j}(t),\Delta_{g_{1},j}(t),\Omega_{g_{1}g_{r},j}(t),\Delta_{g_{r},j}\}$. Notice that having both coupling and detuning allows full control on the Bloch sphere of each individual qubit, allowing \textit{rotational control} \cite{rotational}. For all pulse optimizations in this work, we consider full control over all coupling strengths $\Omega_j$ and detunings $\Delta_j$ available in this system.

\medskip

The Rydberg states $|r\rangle$ are highly excited states that have a passive `always-on' interaction, which is described by a configuration-dependent Hamiltonian $H_V[X]$ \cite{morgado} as a Van der Waals interaction (VdW)  \cite{vdwaals} or a Dipole-Dipole (Dip.) interaction (where $g_1$ needs to be chosen as another Rydberg state), depending on the specific Rydberg states chosen \cite{rydberg1} 
\begin{equation}
\begin{aligned}
H_{V,\text{VdW}}[X]&=\sum_{i=1}^m\sum_{j>i}^m\frac{C_6}{\|x_i-x_j\|^6}|rr\rangle_{ij}\langle rr|_{ij},\\
H_{V,\text{Dip.}}[X]&=\sum_{i=1}^m \sum_{j>i}^m \frac{C_3}{\|x_{i}-x_{j}\|^3}\bigg(|g_1r\rangle_{ij}\langle rg_1|_{ij}+|rg_1\rangle_{ij}\langle g_1r|_{ij}\bigg).
\label{eq:rydbergvdwinteraction}
\end{aligned}
\end{equation}

where $\|x_{i}-x_{j}\|$ is the interatomic distance between atoms $i$ and $j$ and $C_{3,6}$ is an interaction coefficient. The VdW interaction $|rr\rangle_{ij}\langle rr|_{ij}$ shifts the energy level of the doubly excited state scaling with $\|x_i-x_j\|^6$. For close enough atoms, this shift becomes high enough that the doubly excited state becomes unadressable, resulting in a so-called Rydberg-blockade \cite{blockade}. \\

For gg-qubits (ground-ground) we make the identification $|0\rangle=|g_0\rangle,|1\rangle=|g_1\rangle$, and the Rydberg state is used as an auxiliary state interacting via VdW interactions. The state space now becomes that of a qutrit with dimension $d=3^N$. In gr-qubits, $|0\rangle =|g_1\rangle$ and $|1\rangle=|r\rangle$, again interacting with VdW interactions. For this choice of qubit, the Rydberg blockade may cause a part of the computational space to become unreachable. Lastly, we can again consider $|0\rangle= |g_1\rangle$ and $|1\rangle=|r\rangle$ but now with Dipole-Dipole interactions. This configuration would not have the problem of the Rydberg blockade but is experimentally harder to facilitate \cite{morgado}.\\

As also mentioned in the main text, we will take the interaction strength coefficient $C_{3,6}=1$ in arbitrary units. All subsequent pulse strengths and interatomic distances are expressed in terms of this interaction energy coefficient. Note that for other units, time can always be rescaled so that $C_{3,6}=1$ by the Schr\"{o}dinger equation \eqref{eq:hamiltoniaoc}.
\end{document}